\documentclass[pra,twocolumn,showpacs,floatfix]{revtex4-1}
\usepackage{amsmath}
\usepackage{bm}
\usepackage{graphicx}
\usepackage{amssymb}
\usepackage{datetime}
\usepackage{braket}
\usepackage{mathtools}
\usepackage{xcolor}
\usepackage[multidots]{grffile}
\usepackage{rotating}
\usepackage{array}
\usepackage{multirow}
\usepackage{siunitx}
\usepackage{times}
\usepackage{xcolor}
\begin{document}
\newcommand{\figdir}{.}
\newcommand{\figwidth}{.99\columnwidth}
\newcommand{\ffigwidth}{0.4\columnwidth}
\newcommand{\un}{\mathcal{U}}
\newcommand{\cs}{\frac{ d\sigma}{ d\Omega}}
\newcommand{\cstext}{\rm d\sigma / \rm d\Omega}
\newcommand{\csel}{\left.\frac{\rm d\sigma}{\rm d\Omega}\right|_{\rm el}}
\newcommand{\csin}{\left.\frac{\rm d\sigma}{\rm d\Omega}\right|_{\rm inel}}
\newcommand{\bcsin}{\left.\frac{\rm d\sigma}{\rm d\Omega}\right|_{\rm inel}^{\rm Bog}}
\newcommand{\ecsin}{\left.\frac{\rm d\sigma}{\rm d\Omega}\right|_{\rm inel}^{\rm Exact}}
\newcommand{\qtil}{\tilde{q}}
\newcommand{\PreserveBackslash}[1]{\let\temp=\\#1\let\\=\temp}
\let\PBS=\PreserveBackslash

\newcommand{\hint}{\hat{H}_\textrm{int}}
\newcommand{\phisf}{ \ket{\phi_\textrm{sf}}}
\newcommand{\phimi}{ \ket{\phi_\textrm{mi}}}
\newcommand{\miphi}{\bra{\phi_\textrm{mi}}}
\newcommand{\hmu}{K_\textrm{BH}}
\newcommand{\kint}{K_\textrm{0}}
\newcommand{\kinttil}{\tilde{K}_\textrm{0}}
\newcommand{\ktil}{\tilde{K}_\textrm{BH}}
\newcommand{\hV}{V}
\newcommand{\ano}{\hat{a}}
\newcommand{\anod}{\hat{a}^\dagger}
\newcommand{\kin}{\bm{k}_\textrm{in}}
\newcommand{\Ein}{E_\textrm{in}}

\newcommand{\ESF}[1]{\ket{\phi_{#1}}}
\newcommand{\ESC}[2]{\ket{\phi_{#1}^{(#2)}}}
\newcommand{\ESU}[1]{\ket{\mathbf{n}_{#1}}}
\newcommand{\EEFt}[1]{\tilde{E}_{#1}}
\newcommand{\EECt}[2]{\tilde{E}_{#1}^{(#2)}}
\newcommand{\EEUt}[1]{\tilde{\mathcal{E}}_{#1}}
\newcommand{\MEVU}[2]{\braket{\bm\nu_{#1} | \hV | \bm\nu_{#2}}}
\newcommand{\intFF}{\nu}
\newcommand{\nvec}{\mathbf{n}}
\newcommand{\epstil}{\tilde{\varepsilon}}

\title{Matter-wave scattering from strongly interacting bosons in an optical lattice}
\newcommand{\freiburg}{Physikalisches Institut, Albert-Ludwigs Universit\"{a}t Freiburg, Hermann-Herder Stra{\ss}e 3, D-79104, Freiburg, Germany}
\author{Klaus Mayer}
\author{Alberto Rodriguez}
\author{Andreas Buchleitner}
\email[]{a.buchleitner@physik.uni-freiburg.de}
\affiliation{\freiburg}
\date{$Rev: 142 $, compiled \today, \currenttime}
%
\begin{abstract}
We study the scattering of a matter-wave from an interacting system of bosons in an optical lattice, focusing on the strong-interaction regime. Analytical expressions for the many-body scattering cross section are derived from a strong-coupling expansion and a site-decoupling mean-field approximation, and compared to numerically obtained exact results. In the thermodynamic limit, we find a non-vanishing inelastic cross section throughout the Mott insulating regime, which decays quadratically as a function of the boson-boson interaction. 
\end{abstract}

\maketitle

\section{Introduction}
\label{sec-introduction}
In a many-particle system, complexity owed to its multipartite nature arises from two fundamental sources: the indistinguishability of the particles constituting the system and their mutual interactions. In today's ultracold atoms experiments, ensembles of interacting identical particles can be trapped in optical potentials \cite{GreF08,WinS13} and controlled with extreme precision, offering the opportunity to study complex many-body dynamics beyond mere single-particle or mean-field descriptions.  
Simulations of solid-state and strongly-correlated physics have come into reach \cite{LewSAD07,BloDN12,BloDZ08,HunZH13}. Arguably most prominent was the observation of the superfluid to Mott insulator phase transition of bosons in an optical lattice \cite{GreME02}, but also Fermi-Hubbard models, in \cite{Jor2008,Sch2008} and out of equilibrium \cite{Sch2012,Koe2005,Str2010,PonMK06}, have been realised. Currently, the effects of disorder \cite{Bil2008,RoaEF08,Mod10,KonGZM11,Jen2012} and its interplay with interactions \cite{FalLGFI07,Dei2010,PasMW10,SanL2010,GeiWB12,MulG12,GeiBW13,KonMX15,ErrLT14,AlvLL14,SchHB15} are objects of intensive research, as are implementations of 
laser induced gauge-fields for the simulation of quantum magnetism \cite{MicBZ06, Lin2009, Juz2010, Sim2011, Aid2013, Tro2008}.

One challenge that emerges in these systems concerns the extraction of the relevant information. 
While standard time-of-flight (ToF) measurements allow to infer the expansion dynamics of the system \cite{Bil2008}, more elaborate schemes like Bragg-spectroscopy permit to extract, for instance, spectral information \cite{OzeKSD05, Bis11, DuWY10, ErnGK10, Fab2011, For2011}. Whereas ToF-experiments require the destruction of the sample in each run of the measurement, direct imaging of the system can be achieved in-situ by the scattering of light \cite{RouMR13, SykB11, Ye11, DouB11, JacI12, IdzRL00, LakAT09, YeZL13, Bir95, Wei95}, with current experiments reaching single-site resolution \cite{NelLW07, GemZHC09, Kar2009, Bak2009, BAK2010, She2010, Wei2011, CheNOG15}. 

Inspired by neutron scattering \cite{Shu95,CowW71}, the scattering of an atomic matter-wave has recently been put forward as a complementary in-situ imaging technique \cite{SanMH10,HunHB12,HunHC12}. Additionally, matter-waves in combination with cold-atom crystals appear as convenient simulators of low-dimensional solid-state systems \cite{BarA14,BarA14b,LanL14}. 
Experimentally, a BEC has been used as probe of a cloud of bosonic atoms 
in an optical lattice \cite{GadPR12}, with the elastic scattering signal reflecting the crystal structure of the strongly-interacting system, while, in the weakly-interacting regime, inelastic scattering induces single-particle band excitations. 
It has been shown in earlier works that the inelastic scattering signal of the matter-wave probe decays as a function of the strength of the interaction between the target bosons \cite{SanMH10, MayRB14,May2014}. In the regime where this 
interaction strength is too \emph{weak} to deplete the condensate significantly, we have successfully employed Bogoliubov's theory to 
describe analytically the scattering cross section \cite{MayRB14,May2014}. There, we found that the decay of the inelastic signal always emerges linearly 
and is independent of the system size, number of particles, and condensate depletion. 

In this work, we develop an analytical description of the inelastic scattering cross section for \emph{strong} interactions. In particular, we address the question of whether the inelastic cross section vanishes at 
the superfluid to Mott insulator phase transition.
In Sec.~\ref{sec-MBCS}, the stage is set by introducing the cross section of a many-body target. 
In Sec.~\ref{sec-SCE}, analytical expressions for the inelastic cross section for strong interactions are derived from a strong-coupling expansion [Sec.~\ref{sec-SCE-Intro}], and from a mean-field approximation [Sec.~\ref{sec-MFCS}]. In Sec.~\ref{sec-SCE-Results}, we perform a thorough analysis of our findings, where we compare the analytical predictions obtained in Sec.~\ref{sec-SCE} to exact numerical simulations, before we conclude in Sec.~\ref{sec-Concl}. A few rather technical derivations are left to the appendices.

\section{Many-body Scattering Cross Section}
\label{sec-MBCS}
 
\subsection{The Bose-Hubbard model}
The target that we consider is a collection of interacting bosons in the lowest single-particle energy band of a one-dimensional optical lattice with $L$ sites, described by the Bose-Hubbard model \cite{JakBCGZ98}:
\begin{align}\texttt{}
	\hmu &= \sum_{j=1}^{L}\bigg[\frac{U}{2}\hat{n}_j(\hat{n}_j-1) -\mu\hat{n}_j\bigg] -J\sum_{\langle i,j\rangle}\anod_i\ano_j, \label{eq:GK_bhh} 
\end{align}
where $J>0$ is the strength of the nearest-neighbor hopping, given in the latter Hamiltonian by the sum over all pairs $\langle i,j\rangle$ of nearest-neighboring sites. The onsite boson-boson interaction strength $U$ is derived from a Fermi-pseudopotential \cite{Wod91}, and we restrict our considerations to repulsive interactions with $U>0$. In the above grand-canonical description, the chemical potential $\mu$ fixes the average number $N$ of particles. The operators $\ano_j$ and $\anod_j$ are respectively the annihilation and creation operators of a boson on site $j$, obeying bosonic commutation relations,
\begin{equation}
	\big[\ano_i,\anod_j\big] = \delta_{ij},\quad \big[\ano_i,\ano_j\big] = \big[\anod_i,\anod_j\big] = 0,
\end{equation}
and $\hat{n}_j = \anod_j\ano_j$ counts the number of particles on site $j$. The tunneling strength $J$ and the interaction energy $U$ can be controlled by tuning the optical lattice and by means of magnetic Feshbach resonances \cite{TimTHK99,Chi2010}.
Throughout this work, all energies are measured in units of the recoil energy $E_r=\hbar^2k_{\textrm{L}}^2/2M$, where $M$ is the mass of one boson and $k_\textrm{L}$ is the wavenumber of the laser providing the lattice. The natural length scale is given by the lattice constant $d=\pi/k_{\textrm{L}}$.

Depending on the value of $U/J$, the Bose-Hubbard model exhibits two distinct thermodynamic phases \cite{FisWGF89}. In the non-interacting limit where $U/J\rightarrow0$, the ground state of $\hmu$ is a compressible superfluid (SF) state, in which all bosons condense into the same delocalized single-particle Bloch state of the lattice. 
In the opposite limit $U/J\rightarrow\infty$, the bosons localize on the wells of the lattice, and for an integer filling factor 
\begin{equation}
	\nu\equiv \frac{N}{L}\in\mathbb{N}, 
	\label{eq:iff}
\end{equation}
the ground state of $\hmu$ is given by a Fock state with exactly $\nu$ 
bosons on each site, called a Mott insulator (MI). In the thermodynamic limit, the system undergoes a quantum phase transition from the SF to the MI regime for a finite value of $U/J$. The SF phase is characterized by a non-vanishing superfluid fraction, compressibility and on-site number fluctuations, and a vanishing excitation gap. 
In contrast, at the transition point to the MI phase an energy gap opens, 
occupation number fluctuations are suppressed due to the competition between kinetic and interaction energies, the system becomes incompressible and the superfluid fraction vanishes \cite{RotB03}.

\subsection{Scattering from a Bose-Hubbard target}
\begin{figure}
 \centering 
 \includegraphics[width=\figwidth]{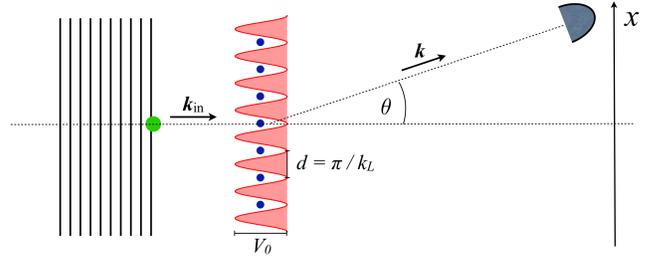}
\caption{(Color online) Scattering setup: A particle of mass $m$ initially in a plane-wave state with momentum $\kin$ is scattered into the angle $\theta$ from a target of atoms (all of which have mass $M$) submerged in a one-dimensional optical lattice with depth $V_0$ and lattice constant $d = \pi/k_L$, where $k_L$ is the laser wavenumber. The asymptotic final state of the probe has momentum $\bm k$. Figure taken from Ref.~\cite{MayRB14}.}
\label{fig:setup}
\end{figure}

We use the formalism presented in Refs.~\cite{MayRB14,May2014}, which for the benefit of the reader we summarize in the following. 
We study the scattering of an atom (probe) of mass $m$ from a Bose-Hubbard system (target), and assume that the optical lattice is transparent to the probe. 
Such a species-selective optical lattice ---based on a species-selective dipole potential \cite{LebT07}--- has already been experimentally realized in a mixture of $^{87}$Rb and $^{41}$K atoms \cite{CatBL09,LamCB10}.
The incoming energy of the probe is considered to be such that no interband excitations of the system can occur, and 
$s$-wave scattering dominates the interaction 
with each target atom. The interaction can therefore be described by the $s$-wave scattering length $a_s$ through a pseudopotential:
\begin{equation}
	V(\bm{r}) = \frac{2\pi\hbar^2}{m}a_s\sum_{\beta=1}^N\delta(\bm{r}-\bm{r}^{(\beta)}),
\end{equation}
where $\bm{r}$ and $\{\bm{r}^{(\beta)}\}_{\beta=1,\dots,N}$ give the positions of the probe and the target atoms, respectively.
The target 
is assumed to be initialized in its ground state $\ket{\phi_0}$, with energy $E_0$, and the probe's asymptotic incoming state is a plane wave $\ket{\kin}$ with wavevector $\kin$, and incoming energy $\Ein=\hbar^2\kin^2/2m$. In Born approximation, the far-field scattering cross section is given by 
\begin{align}
	\frac{1}{a_s^2}\cs =& 
	\sum_{n}\sqrt{1-\frac{E_n-E_0}{\Ein}}\left|\int d{\bm r}\, e^{i{\bm{\kappa}\cdot\bm{r}}}\bra{\phi_n}\hat{n}(\bm{r})\ket{\phi_0}\right|^2,
\label{eq:crosssec_gen}
\end{align}
where $\{\ket{\phi_n}\}$ are the eigenstates of $\hmu$ with corresponding eigenenergies $\{E_n\}$, and $\bm{\kappa}$ is the transferred momentum:
\begin{equation}
	\bm\kappa \equiv \kin-\bm{k},
\end{equation}
with $\bm{k}$ the asymptotic outgoing momentum of the probe. 
We choose the lattice axis along the $\bm{u}_x$-direction, thus $\bm{r}_j$ is given by $\bm{r}_j=x_j \bm{u}_x \equiv j d \bm{u}_x$. For normal incidence of the probe with respect to the lattice axis, the scattering is invariant under rotations around the $x$-axis (neglecting interference terms between the incoming and scattered waves), 
and we can restrict our considerations to the $x$-$y$ plane, as shown in Fig.~\ref{fig:setup}.
The transferred momentum along the direction of the lattice 
\begin{equation}
	 \kappa \equiv \bm{\kappa}\cdot\bm{u}_x,
\end{equation}
obeys
\begin{equation}
	\kappa = \kappa_\text{el}\sqrt{1-\frac{E_n-E_0}{\Ein}},
	\label{eq:kappa}
\end{equation}
where 
\begin{equation}
	\kappa_\text{el} d = -\pi\sin\theta\sqrt{\frac{m}{M}\frac{\Ein}{E_r}},
	\label{eq:kappa_el}
\end{equation}
is the $x$ component of the transferred momentum for elastic scattering. 

The bosonic field operators $\hat{\psi}(\bm{r})$ and $\hat{\psi}^\dagger(\bm{r})$, annihilating and creating a particle at position $\bm{r}$, respectively, define the density operator
\begin{equation}
	\hat{n}(\bm{r}) = \hat{\psi}^\dagger(\bm{r})\hat{\psi}(\bm{r}), \label{eq:dens_wann}
\end{equation}
and they can be expanded in the single-particle Wannier basis:
\begin{equation}
	\hat{\psi}(\bm{r}) = \sum_{j=1}^L\ano_j w(\bm{r}-\bm{r}_j), \label{eq:psi_wann}
\end{equation}
where $w(\bm{r}-\bm{r}_j)$ is the Wannier function describing a particle localized around the position $\bm{r}_j$ of the $j$-th lattice site \cite{Wan37,Koh59}. 
From expansion \eqref{eq:psi_wann} the density operator 
is split into diagonal and off-diagonal contributions in the Wannier basis, the latter being proportional to the overlap of Wannier functions centered at different sites. We consider a deep lattice of depth $V_0=15E_r$, for which this overlap can safely be neglected 
\cite{MayRB14}. 
In this \emph{diagonal approximation}, the inelastic cross section (i.e.~when the target is left in an excited state) reads 
\begin{align}
	\frac{1}{a_s^2}\frac{d\sigma}{d\Omega}\bigg|_\textrm{inel} =&  \sum_{n\neq0}
	\sqrt{1-\frac{E_n-E_0}{E_\textrm{in}}} \notag\\
	&\times \bigg| \sum_{j=1}^{L} e^{i\kappa x_j}\braket{\phi_n | \hat{n}_j | \phi_0}\bigg|^2\big|W(\kappa)\big|^2,
	\label{eq:CS_inel_main_WANN}
\end{align}
where $W(\kappa)$ is the form factor of a unit cell of the lattice:
\begin{equation}
	W(\kappa) = \int dx\, e^{i\kappa x}\, \big|w(x)\big|^2.\label{eq:form_factor}
\end{equation}

The inelastic cross section bears a clear signature of the nature of the ground state. 
In the SF regime, fluctuations of the on-site occupation numbers give rise to a non-vanishing inelastic cross section, which can be described analytically both at $U=0$ (from the exact solution of $\hmu$) \cite{SanMH10}, as well as in the regime of weak-depletion of the macroscopically occupied single-particle ground state \cite{MayRB14}, using Bogoliubov's theory. 
The inelastic cross section decays as a function of $U/J$, and in the MI limit ($U/J\rightarrow\infty$), it vanishes exactly in the diagonal approximation 
(see Table I in Ref.~\cite{MayRB14}), as the ground state becomes an eigenstate of the on-site number operator $\hat{n}_j$. 

In this work, we want to reveal how the inelastic cross section depends on the interaction in the regime of strong, but finite interactions. To this aim, we employ a strong-coupling expansion (SCE) of Hamiltonian \eqref{eq:GK_bhh}, in which a perturbative series of its eigenstates and energies is constructed by regarding the tunneling operator as a perturbation to the pure interaction Hamiltonian \cite{FreM94,FreM96,EjiFGM12}. Moreover, we also show how 
a site-decoupling mean-field approach fails to describe the behavior of the cross section. 

\section{Derivation of the inelastic cross section in the strongly-interacting regime}
\label{sec-SCE}
\subsection{Strong-Coupling expansion}
\label{sec-SCE-Intro}
In order to set up a perturbative expansion, the grand-canonical Bose-Hubbard Hamiltonian [Eq.~\eqref{eq:GK_bhh}] is split into the unperturbed on-site part,
\begin{equation}
	\kinttil = \sum_{j=1}^L\left[\frac{1}{2}\hat{n}_j\,(\hat{n}_j-1)-\tilde{\mu}\hat{n}_j\right],\label{eq:kinttil}
\end{equation}
and a perturbation given by
\begin{equation}
	\hV = \sum_{j=1}^{L}\left(\anod_{j}\ano_{j+1}+\anod_{j+1}\ano_{j}\right),
\end{equation}
with all energies in units of $U$, indicated by the tilde notation, i.e.~$\tilde{\mu} \equiv \mu/U$. 
For small values of the dimensionless tunneling strength $\tilde{J} \equiv J/U$, the eigenstates and eigenenergies of the full Hamiltonian
\begin{equation}
	\ktil = \kinttil  - \tilde{J}\,\hV\label{eq:HBH_tilde}
\end{equation}
can be obtained as a power series in $\tilde{J}$. 
Notice that $\ktil$, $\kinttil$ and $\hV$, all commute with the total number operator $\hat{N}=\sum_{j=1}^L\hat{n}_j$, and therefore all energy eigenstates can be chosen to have a well-defined number of bosons.  

The eigenstates of $\kinttil$ can be taken 
as eigenstates of the on-site number operators $\hat{n}_j$,
\begin{align}
    \ket{\nvec}&\equiv\ket{n_1,\dots,n_L}, \\
	\hat{n}_j\ket{\nvec} &= n_j\ket{\nvec},\,n_j \in \mathbb{N}_0,
\end{align}
with unperturbed energy $\EEUt{\mathbf{n}}$:
\begin{align}
	\kinttil\ESU{} &= \EEUt{\nvec}\ESU{},\\
	\EEUt{\nvec}&=\sum_{j=1}^L\left[\frac{1}{2}n_j(n_j-1)-\tilde{\mu}n_j\right].
\end{align}
For integer filling factor $\intFF$ [Eq.~\eqref{eq:iff}], the ground state of $\kinttil$,
\begin{equation}
	\ket{\intFF,\dots,\intFF}\equiv\phimi, 
\end{equation}
is non-degenerate. It corresponds to $\intFF$ particles localized on each site, and has unperturbed energy
\begin{equation}
	\EEUt{\textrm{mi}} = L\left[\frac{1}{2}\intFF(\intFF-1)-\tilde{\mu}\intFF\right].
\end{equation}
We write the perturbation series of the ground state $\ESF{0}$ of $\ktil$ as 
\begin{align}
	\ktil\ESF{0} &= \EEFt{0}\ESF{0},\\
	\ESF{0} &= \phimi-\tilde{J}\ESC{0}{1} + \mathcal{O}(\tilde{J}^2),\label{eq:gs_expansion}\\
	\EEFt{0} & =\EEUt{\textrm{mi}} -\tilde{J}\EECt{0}{1}+\tilde{J}^2\EECt{0}{2} + \mathcal{O}(\tilde{J}^3).
\end{align}
To first order in the tunneling, the correction to this state is readily obtained from non-degenerate perturbation theory (PT):
\begin{equation}
	\ESC{0}{1} = \sum_{\ESU{}\neq\phimi}\frac{ \braket{\nvec | \hV | \phi_\textrm{mi} }}{\EEUt{\textrm{mi}}-\EEUt{\nvec}}\ESU{} = -\hV \phimi. 
\end{equation}
Hence, the first correction is a superposition of states with one additional particle-hole (PH) pair, i.e.~of states of the form $\ESU{}=\ket{\intFF,\dots,\intFF\pm1,\intFF\mp1,\dots,\intFF}$, with excitation energy
\begin{equation}
	\EEUt{\nvec}-\EEUt{\textrm{mi}} = 1.
\end{equation}	
In the ground state, the first-order correction to the energy $\EECt{0}{1}\equiv \miphi\hV\phimi$ vanishes, since $\hV$ only connects states that differ by one PH pair. 
More generally, all odd-order corrections to the ground-state energy can be shown to vanish. To third order, the perturbed energy of the ground state is obtained  straightforwardly \cite{FreM94,FreM96}:
\begin{equation}
	\EEFt{0} = \EEUt{\textrm{mi}}  -2 L\intFF(\intFF+1) \tilde{J}^2 + \mathcal{O}(\tilde{J}^4),
	\label{eq:Mott_energy_2ndorder}
\end{equation}
Symbolical expansions for ground-state properties can be obtained on a computer within non-degenerate perturbation theory to very high orders \cite{EjiFGM12,DamZ06}.

The inelastic cross section [Eq.~\eqref{eq:CS_inel_main_WANN}] is governed by the matrix elements
\begin{equation} 
	\braket{\phi_n | \hat{n}_j | \phi_0},
	\label{eq:ME}
\end{equation}
which are
non-zero only if the states $\ket{\phi_n}$ and $\ket{\phi_0}$ have 
the same number of particles.
Therefore, the SCE of \eqref{eq:ME} requires also the perturbative series of the fixed-density excitations of $\kinttil$, which can be characterized by the number of PH pairs. 
Given that the tunneling operator changes the number of PH pairs by one in every subsequent term of the perturbation series of a state, the desired order in $\tilde{J}$ for the
matrix elements dictates which manifolds of excitations need to be taken into account: For the expansions of \eqref{eq:ME} to order $r$, the perturbative series to that order of all $\kinttil$-excitations including up to $r$ PH pairs, and the ground state, must be considered.
To first order, only one-PH states of the form 
\begin{equation}
	\ket{s,s+l} = \frac{\anod_s\ano_{s+l}}{\sqrt{\intFF(\intFF+1)}}\phimi, \quad\begin{tabular}{l} $s=1,\dots,L$,\\ $l=1,\dots,L-1$,\end{tabular}
	\label{eq:PH_states_degen}
\end{equation}
contribute, where $s$ and $s+l$ denote the position of the particle and the hole, respectively, and periodic boundary conditions are used \cite{EjiFGM12}. The restriction on the distance $l$ enforces that particle and hole be on different sites.
The states~\eqref{eq:PH_states_degen} span an $L(L-1)$-fold degenerate subspace $\mathcal{D}$ with unperturbed excitation energy
\begin{equation}
	\EEUt{\textrm{ph}}-\EEUt{\textrm{mi}}=1.\label{eq:PH_unp_excen}
\end{equation}
Thus, we need to employ degenerate PT, and find states with a well-defined limit as $\tilde{J}\rightarrow0$. The latter are obtained by diagonalizing 
the perturbation $\hV$ in the degenerate subspace $\mathcal{D}$. 
Translational invariance of $\hmu$ allows to separate $\mathcal{D}$ into $L$ manifolds characterized by different well-defined center-of-mass (CoM) quasimomenta $q$ \cite{FehSW,May2014,EjiFGM12}:
\begin{equation}
	\ket{q,l} = \sqrt{\frac{1}{L}}\sum_{s=1}^{L}e^{iqsd}\ket{s,s+l},\quad 
	\begin{tabular}{l} 
	$q=2\pi j/Ld$,\\ $ j=0,\dots,L-1.$\end{tabular}\label{eq:q_ph_states}
\end{equation}
In this basis, the CoM degree of freedom decouples and the perturbation becomes block-diagonal \cite{GolMG09}:
\begin{align}
	\braket{q', l' | \hV | q, l} = \delta_{q,q'}\left[T_q^*\delta_{l',l-1} + T_q\delta_{l',l+1}\right]\label{eq:PH_states_tridiag},
\end{align}
with the effective (complex) tunneling parameter
\begin{equation}
	T_q =(\intFF+1)e^{iqd} +\intFF. \label{eq:compl_tunn}
\end{equation}
The matrix \eqref{eq:PH_states_tridiag} 
describes the problem of a particle on an open chain with hard-wall boundary conditions at sites $L$ and a fictitious site $0$ (since $l$ has to be different from both, $0$ and $L$). Diagonalization is 
achieved by the states 
\begin{equation}
	\ket{q, \varkappa} = \sqrt{\frac{2}{L}}\sum_{l=1}^{L-1}\sin(\varkappa d\,l)e^{i\zeta_ql}\ket{q,l},\quad 
	\begin{tabular}{l}
	$\varkappa = \pi j'/Ld$, \\ $j'=1,\dots,L-1,$\end{tabular}\label{eq:qk_states}
\end{equation}
where we note that $\varkappa$ can be interpreted as the quasimomentum connected to the PH relative coordinate. 
The phase $\zeta_q$ is the argument of the complex tunneling parameter $T_q$, and given by
\begin{equation}
	\tan{\zeta_q} = \frac{(\intFF+1)\sin(qd)}{\intFF+(\intFF+1)\cos(qd)}.\label{eq:zeta_q}
\end{equation}
The eigenvalue of the perturbation (restricted to $\mathcal{D}$) is 
\begin{equation}
	\EECt{q,\varkappa}{1} = 2 \cos(\varkappa d)\sqrt{1+4\intFF(\intFF+1)\cos^2(qd/2)},
	\label{eq:PH_disp_rel}
\end{equation}
which
gives the first order energy correction of the eigenstates $\ket{\phi_{q,\varkappa}}$ of $\hmu$ whose ($\tilde{J}=0$)-limit are the one-PH states $\ket{q,\varkappa}$:
\begin{align}
  \ket{\phi_{q,\varkappa}} &= \ket{q,\varkappa}-\tilde{J} \ket{\phi_{q,\varkappa}^{(1)}} + \mathcal{O}(\tilde{J}^2), \label{eq:ph_states_expansion} \\
  \EEFt{q,\varkappa} &= \EEUt{\textrm{ph}} - \tilde{J}\EECt{q,\varkappa}{1}  + \tilde{J}^2\EECt{q,\varkappa}{2} +\mathcal{O}(\tilde{J}^3). \label{eq:PH_enegy}
\end{align}
Note that to first order
the energy is still degenerate in the center-of-mass quasimomentum $q$. This degeneracy, however, plays no role for the perturbative expansion,
since, as mentioned above, the perturbation is translationally invariant and thus never couples states with different values of $q$. One can then proceed as if the degeneracy were fully lifted \cite{May2014}. 
The second order correction to the energy is given in Appendix \ref{app:SCEformulas}. 
The first-order correction to the state can be written as
\begin{equation}
	\ket{\phi_{q,\varkappa}^{(1)}} = \braket{ \phi_\textrm{mi}| \hV | q,\varkappa }\phimi  + \ket{v_{\mathcal{C}}} +  \ket{v_{\mathcal{D}}}.
	\label{eq:qk_firstorder}
\end{equation}
Here, the first term corresponds to the ground-state contribution, $\ket{v_{\mathcal{C}}}$ to a state orthogonal to both, $\phimi$ and $\mathcal{D}$, and $\ket{v_{\mathcal{D}}}$ is the contribution of all states from within $\mathcal{D}$ with $\varkappa'\neq\varkappa$. The explicit expressions to obtain $\ket{v_{\mathcal{C}}}$ and $\ket{v_{\mathcal{D}}}$ are given in Appendix~\ref{app:SCEformulas}. 
However, for a first-order expansion of the matrix element~\eqref{eq:ME}, only the ground-state contribution in Eq.~\eqref{eq:qk_firstorder} is necessary, since $\ket{v_{\mathcal{C}}}$ and $\ket{v_{\mathcal{D}}}$ couple to $\ESF{0}$ [Eq.~\eqref{eq:gs_expansion}] only to higher orders in $\tilde{J}$.

We would like to emphasize that SCE provides a rather accurate description of the phase diagram of the Bose-Hubbard model, which can be 
obtained from the energy gap between the ground state and the lowest non-number-conserving excitations of the system (i.e.~adding/removing one particle) \cite{FreM94,FreM96,EjiFGM12}. The resulting equation for the critical point of the SF-MI phase transition at fixed density is given in Appendix \ref{app:SCEformulas}. 

\subsection{Inelastic cross section from SCE}
The leading contribution in $\tilde{J}$ of the inelastic cross section [Eq.~\eqref{eq:CS_inel_main_WANN}] is obtained by considering all matrix elements \eqref{eq:ME} with a non-vanishing first order term. As explained above, only the manifold of eigenstates $\ket{\phi_{q,\varkappa}}$ needs to be taken into account. 
From the expansions 
\eqref{eq:gs_expansion} and 
\eqref{eq:ph_states_expansion}, the matrix element of the onsite-density operator is found to be 
\begin{align}
	\braket{\phi_{q,\varkappa}| \hat{n}_j | \phi_0 } &= \tilde{J} \braket{q,\varkappa| [\hat{n}_j,V] | \phi_\text{mi} }+ \mathcal{O}(\tilde{J}^2) \notag\\
	&=\tilde{J}\frac{\sqrt{2\intFF(\intFF+1)}}{L}M(q,\varkappa)e^{-iqjd}+ \mathcal{O}(\tilde{J}^2),
	\label{eq:SCE_ME_1st}  
\end{align}
with $M(q,\varkappa)$ given by
\begin{align}
	M(q,\varkappa)=& -2i\sin(\varkappa d)\sin(qd/2) \notag\\ 
	&\times \left[ e^{i(q/2-\zeta_q)}+ \cos(\varkappa dL)e^{-i(q/2+\zeta_q(L-1))}\right].
	\label{eq:def_M}
\end{align}
Inserting Eqs.~\eqref{eq:Mott_energy_2ndorder}, \eqref{eq:PH_enegy} and \eqref{eq:SCE_ME_1st} into Eq.~\eqref{eq:CS_inel_main_WANN}, we 
obtain an expression for the inelastic scattering cross section, valid for large values of the interaction strength:
\begin{align}
	\frac{1}{Na_s^2}\frac{d\sigma}{d\Omega}\bigg|_\textrm{inel}^\textrm{SCE} =& \left[\frac{J}{U}\right]^2 \frac{2(\intFF+1)}{L^3} 
	\sum_{q,\varkappa} \sqrt{1-\frac{U(1-\tilde{J}\EECt{q,\varkappa}{1})}{E_{\textrm{in}}}}\nonumber\\
	&\times\Big|M(q,\varkappa)\Sigma(\kappa-q)W(\kappa)\Big|^2,
	\label{eq:SCE_CS_fin}
\end{align}
where the excitation energy gap has been evaluated to first order in $\tilde{J}$ only, and 
we normalized to the number of bosons $N$, the scale of the inelastic cross section in the non-interacting limit \cite{MayRB14}.
The function $\Sigma(\kappa-q)\equiv\sum_{j=1}^Le^{i(\kappa-q)x_j}$ represents the interference of elementary waves emanating from each scattering centre, 
\begin{align}
	|\Sigma(\kappa-q)|^2 = \frac{\sin^2\big((\kappa-q)dL/2\big)}{\sin^2\big((\kappa-q)d/2\big)}.
\end{align}
Notice that, although not indicated explicitly, the transferred momentum $\kappa$ [Eq.~\eqref{eq:kappa}] depends on the transferred energy, and thus on the quasimomenta $q$ and $\varkappa$ via
\begin{equation}
	\kappa = \kappa_{\textrm{el}}\sqrt{1-\frac{U(1-\tilde{J}\EECt{q,\varkappa}{1})}{\Ein}}.
	\label{eq:kappa_SCE}
\end{equation}
Equation~\eqref{eq:SCE_CS_fin} shows that, for large values of the interaction, the inelastic cross section decays quadratically with $U/J$, 
in agreement with the behavior of the dynamic structure factor of a Bose-Hubbard system \cite{GolMG09}. 
The dynamic structure factor is intimately linked to the scattering cross section, being the response of the system to a density perturbation \cite{MenKPS03, RouMR13}.

\subsection{Site-decoupling mean-field description}
An effective 
Hamiltonian results from~\eqref{eq:GK_bhh} when the tunneling (the coupling of a given site to its neighbors) is treated in terms of a mean-field (MF) coupling \cite{FisWGF89,Sto09,Sac99,GeoG2010}, i.e.~when the full hopping term in \eqref{eq:GK_bhh} is replaced by 
\begin{equation}
	-J\sum_{\langle i,j\rangle}\anod_i\ano_j \longrightarrow -\sum_{j=1}^{L}\big(\lambda_j\anod_j+\lambda_j^*\ano_j\big),
\end{equation}
where in one dimension the mean-field coupling strength reads 
\begin{equation}
	\lambda_j = J\Big(\langle\ano_{j+1}\rangle + \langle\ano_{j-1}\rangle\Big),
\end{equation}
and the expectation values are taken with respect to the ground state of the system. 
In a translationally invariant lattice, and due to the invariance of $\hmu$ under a global U(1)-transformation, the above quantity can be chosen to be real and constant for all lattice sites.
Then, the MF coupling strength $\lambda\equiv\lambda_j$ is fixed by the self-consistency relation
\begin{equation}
	\lambda = 2J \langle\ano_j\rangle, \label{eq:lambda_SelfCons}
\end{equation}
since 
 the expectation value itself depends on $\lambda$.
The resulting effective Hamiltonian is a sum of single-site operators:
\begin{equation}
	\tilde{H}_{\textrm{eff}} = \sum_{j=1}^{L} \tilde{h}^{(j)} \equiv \sum_{j=1}^{L}\left[ \tilde{h}_{0}^{(j)} + \tilde{\lambda} v^{(j)}\right],
	\label{eq:MFT_full_Ham}
\end{equation}
where 
\begin{align}
	\tilde{h}_{0}^{(j)} &= \frac{1}{2}\hat{n}_j(\hat{n}_j-1)-\tilde{\mu}\,\hat{n}_j,\label{eq:MFT_SS_Ham}\\
	v^{(j)} &= -(\anod_j+\ano_j),\label{eq:MFT_SS_pert}
\end{align}
and $\tilde{\lambda} \equiv \lambda/U$.
Numerically, it is found iteratively that $\tilde{\lambda}$ is the mean-field order parameter of the SF-MI phase transition \cite{FisWGF89,Sto09,Sac99,GeoG2010}: it is zero in the MI phase, and takes on non-zero values continuously as the phase boundary into the SF regime is crossed. Therefore, in the vicinity of the critical point, $\tilde{\lambda}$ is small, which motivates a perturbative expansion of the eigenstates and eigenenergies of $\tilde{H}_{\textrm{eff}}$ in $\tilde{\lambda}$. 
The fact that $\tilde{H}_{\textrm{eff}}$ is a sum of single-site operators allows us to reduce the calculation to the single site Hamiltonian 
$\tilde{h}\equiv\tilde{h}^{(j)}$. 

The single-site eigenstates $\ket{\nu}$ of $\tilde{h}_{0}$ are characterized by the integer number of bosons $\nu$, and have unperturbed energies 
\begin{equation}
	\tilde{\varepsilon}(\intFF) = \frac{1}{2}\intFF(\intFF-1)-\tilde{\mu} \intFF,\label{eq:MFT_unpert_energ}
\end{equation}
in units of the interaction energy $U$. In order for $\ket{\nu}$ to be the state of lowest energy, the chemical potential has to satisfy
\begin{equation}
	\intFF-1<\tilde{\mu}<\intFF,\quad \nu\in\mathbb{N}\label{eq:mft_mu_restr}.
\end{equation}
The perturbative expansion of a generic eigenstate $\ket{\xi_\intFF}$ of $\tilde{h}$ reads 
\begin{equation}
	\ket{\xi_\intFF} = \ket{\intFF} + \tilde{\lambda}\ket{\xi_\intFF^{(1)}} +\tilde{\lambda}^2\ket{\xi_\intFF^{(2)}} + \mathcal{O}(\tilde{\lambda}^3),\label{eq:MFT_singsite_exp1}
\end{equation}
where the 
corrections follow from standard non-degenerate perturbation theory: 
\begin{align}
	\ket{\xi_\intFF^{(1)}} =& - c_\intFF^{(-1)} \ket{\intFF-1} -c_\intFF^{(1)} \ket{\intFF+1},  \label{eq:MFT_singsite_exp2} \\
	\ket{\xi_\intFF^{(2)}} =&  c_\intFF^{(-1)}c_\intFF^{(-2)} \ket{\intFF-2} +c_\intFF^{(1)}c_\intFF^{(2)}\ket{\intFF+2} \\
	  &- \frac{1}{2}\left( [c_\intFF^{(-1)}]^2 + [c_\intFF^{(1)}]^2 \right)\ket{\intFF},
\end{align}
in terms of the coefficients 
\begin{equation}
	c_\intFF^{(k)} = \frac{\sqrt{\intFF+k+[1-\operatorname{sgn}k]/2}}{\epstil(\intFF)-\epstil(\intFF+k)}, \quad k\neq0. 
	\label{eq:coeff_c_n}
\end{equation}
Similarly, the perturbative expansion of the energy $\tilde{\epsilon}_\intFF$ of $\ket{\xi_\intFF}$ reads 
\begin{align}
  \tilde{\epsilon}_\intFF &= \epstil(\intFF) + \left( \sqrt{\intFF}\,c_\intFF^{(-1)} + \sqrt{\intFF+1}\,c_\intFF^{(1)} \right) \tilde{\lambda}^2+\mathcal{O}(\tilde{\lambda}^4). 
  \label{eq:MF_energyseries}
\end{align}

Using the second-order expansion of $\ket{\xi_\intFF}$, 
the value of the MF coupling strength $\tilde{\lambda}$ in the SF phase is estimated 
from the self-consistency condition \eqref{eq:lambda_SelfCons} \cite{May2014}: 
\begin{align}
	\tilde{\lambda}^2 =& -\frac{1}{B}\left(A+\frac{1}{2\tilde{J}}\right), \label{eq:MFT_lambda2}\\
	A =& \sqrt{\intFF}\,c_\intFF^{(-1)}+\sqrt{\intFF+1}\,c_{\intFF}^{(1)},\\
	B =& \sqrt{\intFF+2}\,[c_{\intFF}^{(1)}]^2 c_{\intFF}^{(2)} +  \sqrt{\intFF-1}\,[c_{\intFF}^{(-1)}]^2 c_{\intFF}^{(-2)} \notag\\ 
	&- \frac{1}{2}A\left( [c_\intFF^{(-1)}]^2 + [c_\intFF^{(1)}]^2 \right).
\end{align}
\subsection{Inelastic cross section from the MF description}
\label{sec-MFCS}
Due to the separability of $\tilde{H}_{\textrm{eff}}$ [Eq.\eqref{eq:MFT_full_Ham}], the calculation of the matrix elements \eqref{eq:ME} in the inelastic cross section reduces to the single-site level. 
We therefore need to evaluate the matrix element $\braket{\xi_{\intFF'} | \hat{n} | \xi_{\intFF}}$ 
for the ground and excited states $\ket{\xi_{\intFF}}$ and $\ket{\xi_{\intFF'}}$, respectively, of the single-site Hamiltonian $\tilde{h}$. 
From Eq.~\eqref{eq:MFT_singsite_exp2} one can see that, to first order in $\tilde{\lambda}$, 
only excited states with $\nu'=\nu\pm1$ must be considered:
\begin{align}
	\Braket{\xi_{\intFF+1} | \hat{n} | \xi_{\intFF}} &=-\tilde{\lambda}\,c_{\intFF}^{(1)} +\mathcal{O}(\tilde{\lambda}^2), \nonumber\\
	\Braket{\xi_{\intFF-1} | \hat{n} | \xi_{\intFF}} &=\tilde{\lambda}\,c_{\intFF}^{(-1)} +\mathcal{O}(\tilde{\lambda}^2).\label{eq:MF_ME}
\end{align}
The contributing many-site excited states are of the form $\ket{\xi_{\nu\pm1}}_j\otimes_{i\neq j}\ket{\xi_\intFF}_i$, for $j=1,\ldots,L$. Therefore, to leading order in $\tilde{\lambda}$, the inelastic cross section [Eq.~\eqref{eq:CS_inel_main_WANN}] is given by $L$ times the single-site contribution \cite{May2014}. 
To first order in $\tilde{\lambda}$, the energy gap to the contributing excited states is given by [see Eq.~\eqref{eq:MF_energyseries}]
\begin{equation}
	\tilde{\Delta}^{\pm}_\intFF \equiv \epstil(\intFF\pm1)-\epstil(\intFF). 
\end{equation}

Finally, to derive the MF cross section, the value of the chemical potential $\tilde{\mu}$ needs to be specified. 
We aim at a description of the scattering cross section in the vicinity of the \emph{fixed-density} (FD) SF-MI transition. The corresponding value of the chemical potential can be obtained from the MF phase diagram of the Bose-Hubbard model \cite{FisWGF89} (for details see, e.g., Ref.~\cite{May2014}):
\begin{equation}
	\tilde{\mu}_\textrm{FD}(\intFF) = \sqrt{\intFF(\intFF+1)}-1.
\end{equation}
Putting everything together yields the MF inelastic scattering cross section to leading order in $\tilde{\lambda}$, in the strong interaction regime: 
\begin{align}
	\frac{1}{N a_s^2}\frac{d\sigma}{d\Omega}\bigg|_\textrm{inel}^\textrm{MF} =& \frac{\tilde{\lambda}^2}{\intFF} \sum_{\sigma=\pm} \sqrt{1-\frac{U\tilde{\Delta}^\sigma_\intFF}{E_\textrm{in}}}\,[c_\intFF^{(\sigma 1)}]^2 |W(\kappa^\sigma)|^2, 
	\label{eq:MFT_inel_CS}
\end{align}
with the transferred momenta
\begin{equation}
	\kappa^{\pm} = \kappa_\textrm{el}\sqrt{1-U\tilde{\Delta}^\pm_\intFF/\Ein},
\end{equation}
and it must be emphasized that $\tilde{\lambda}^2=0$ for $\tilde{J}\leqslant\tilde{J}_c^\text{MF}$, and $\tilde{\lambda}^2$ is given by Eq.~\eqref{eq:MFT_lambda2} for 
 $\tilde{J}>\tilde{J}_c^\text{MF}$, where 
\begin{equation}
  \tilde{J}_c^\text{MF} = 1/2 + \intFF - \sqrt{\intFF (\intFF+1)}
  \label{eq:MF_critpoint}
\end{equation}  
corresponds to the MF estimate of the FD critical point. Hence, according to the MF description, the inelastic cross section should decrease linearly with $U/J$ in the vicinity of the transition when approached from the SF phase, vanish at the critical point and remain zero in the whole Mott phase. 
As one goes deeper into the SF phase (and thus the value of $\tilde{\lambda}$ increases), a higher order expansion of the single-site states would be necessary to estimate correctly the dependence of $\tilde{\lambda}$ on $U/J$, which will include non-linear terms. 

\section{Analysis of the inelastic cross section}
\label{sec-SCE-Results}
In order to gauge the quality of the approximate analytical expressions \eqref{eq:SCE_CS_fin} and \eqref{eq:MFT_inel_CS} for the inelastic cross section, we calculate numerically the exact cross section, Eq.~\eqref{eq:CS_inel_main_WANN}, via exact numerical diagonalization of $\hmu$ in a system with fixed integer density.
The Wannier functions are approximated by Gaussians, 
corresponding to the ground-state wave function of a harmonic approximation of each optical potential well, leading to a Gaussian form factor [Eq.~\eqref{eq:form_factor}], 
\begin{align}
	W(\kappa)=e^{-\frac{(\kappa d)^2}{4\pi^2\sqrt{V_0/E_r}}}.
\end{align}
This is a good and common description for deep enough optical lattices, and valid for our choice of the lattice depth $V_0=15 E_r$, which corresponds to the tunneling strength $J=6.5\times10^{-3}E_r$, and gives rise to a gap to the second band of $6.28 E_r$  \cite{MayRB14}. 

\subsection{MF expression}
\begin{figure}
\centering
	\includegraphics[width=\figwidth]{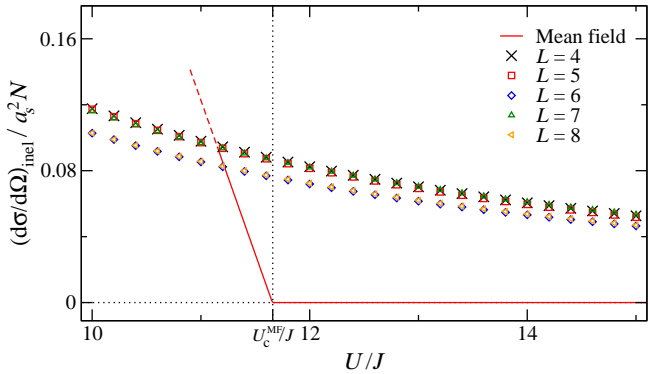}
	\caption{(Color online) Inelastic scattering cross section in the strong-interaction regime for unit filling factor $\intFF=1$. The symbols show exact numerical results for different system sizes, while the red solid line corresponds to the analytical formula \eqref{eq:MFT_inel_CS}, obtained from MFT (only in the region where $\tilde{\lambda}\leqslant0.1$). The latter predicts the vanishing of the inelastic cross section at the critical interaction strength $U^{\textrm{MF}}_\textrm{c} = 11.66J$, indicated by a vertical dotted line. Relevant parameters are $\Ein=2E_r$, $m=M$, and $\theta = 0.99$.}
	\label{fig:MFCS_L8B8}
\end{figure}

Figure \ref{fig:MFCS_L8B8} shows the mean-field prediction [Eq.~\eqref{eq:MFT_inel_CS}] in comparison to exact numerical results for different system sizes at unit filling. Clearly, the mean-field approximation fails to describe the behavior of the inelastic cross section.
As discussed in Sec.~\ref{sec-MFCS}, the MF cross section exactly vanishes in the whole Mott phase up to the estimated critical point $\tilde{J}^\text{MF}$ [Eq.~\eqref{eq:MF_critpoint}]. 
Comparison to numerical results, however, show that---at least for finite system size--- the inelastic cross section is non-zero in this region. 
For the system sizes accessible in our numerical simulations, no trend towards the $L$-independent MF result \eqref{eq:MFT_inel_CS} can be observed.

A quantitative disagreement between the simulations and the MF result should be expected, given the known poor performance of MF in one dimension. More elaborate mean-field like approaches (e.g., a dynamical MF approach \cite{AmiP98}) might provide a better description of the inelastic cross section. However, the discrepancy found here clearly demonstrates that the site-decoupling MF formalism does not yield a qualitatively correct result in one dimension, if employed in the above straightforward manner.

\subsection{SCE expression}
In contrast to the simplistic site-separability of the eigenstates of the system enforced by the MF approach for all values of $U/J$, the SCE of Sec.~\ref{sec-SCE-Intro} 
describes the eigenstates of $\hmu$ for $\tilde{J}\neq0$ in terms of superpositions of Fock states with different occupation distributions, which are in general non-separable in the site basis.

\begin{figure}
\centering
	\includegraphics[width=\figwidth]{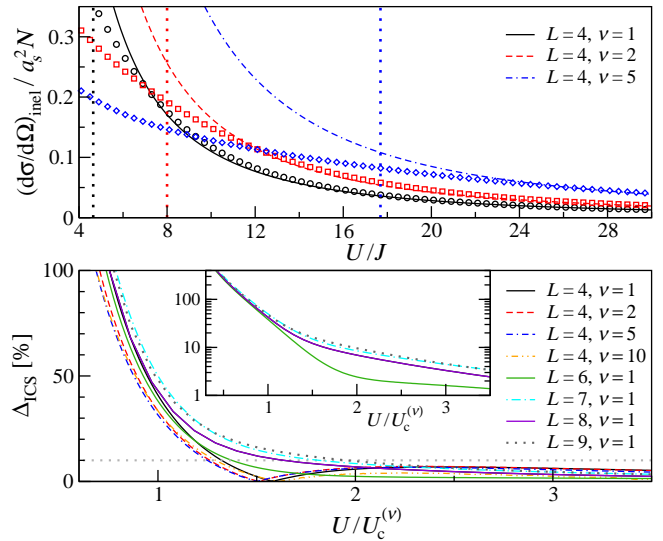}
	\caption{(Color online) Upper panel: Inelastic scattering cross section for a system with $L=4$ sites and different bosonic densities $\nu$. For large enough  
	interaction $U/J$, the SCE formula \eqref{eq:SCE_CS_fin} (lines) correctly describes the exact numerical data (symbols). 
	Vertical dotted lines indicate the critical value $U_\textrm{c}^{(\intFF)}$ of the SF-MI phase transition for each $\nu$ ($\nu =1,2,5$ from left to right), as predicted from third order SCE (see Appendix \ref{app:SCEformulas}). 
	Lower panel: Relative difference $\Delta_\text{ICS}$ [Eq.~\eqref{eq:rel_diff}] between the analytical approximation and the exact numerics, versus the interaction strength in units of the respective critical interaction, for different system sizes and densities. The horizontal dotted line indicates a deviation of $10\%$. 
	The non-monotonicity of $\Delta_\text{ICS}$ visible for $L=4$ is a finite-size effect due to a crossing of the analytical approximation and the exact numerical data.
	The inset shows data for $L\geqslant 6$ 
	on a semi-logarithmic scale. Relevant parameters for both panels are $\Ein=2E_r$, $m=M$, $\theta = 0.99$.} 
	\label{fig:SCECS_L4_diff_n_v2}
\end{figure}

The SCE treatment leads to the $L$-dependent expression \eqref{eq:SCE_CS_fin} for the inelastic cross section, which is predicted to decay quadratically for large $U/J$, and hence to be finite ---at least for finite system size--- inside the Mott phase.
This result is confirmed in the upper panel of Fig.~\ref{fig:SCECS_L4_diff_n_v2}, where Eq.~\eqref{eq:SCE_CS_fin} is compared to the numerically calculated exact cross section for a system size of $L=4$ sites and different filling factors $\intFF$. 
It is however apparent that the higher the density, the larger the interaction has to be in order for the analytical approximation to be valid. 
This is correlated with the fact that for higher densities the critical point of the SF-MI transition is shifted towards stronger interactions \cite{FisWGF89} (in the upper panel of Fig.~\ref{fig:SCECS_L4_diff_n_v2}, the corresponding positions of the critical points $\tilde{J}_c^{(\nu)}$, estimated from SCE as described in Appendix \ref{app:SCEformulas}, are marked by vertical dashed lines). This behavior seems to indicate that the parameter controlling the validity of the perturbative expansion is not $\tilde{J}$ but rather the ratio $\tilde{J}/\tilde{J}_c^{(\nu)}$. 
This conjecture is further supported 
by the lower panel of Fig.~\ref{fig:SCECS_L4_diff_n_v2}, which shows the relative difference
\begin{equation}
      \Delta_\text{ICS}
	=\frac{\Big|\frac{d\sigma}{d\Omega}\big|_\textrm{inel}^\textrm{exact} - \frac{d\sigma}{d\Omega}\big|_\textrm{inel}^\textrm{SCE}\Big|}{\frac{d\sigma}{d\Omega}\big|_\textrm{inel}^\textrm{exact}}\label{eq:rel_diff}
\end{equation}
between the exact inelastic cross section [Eq.~\eqref{eq:CS_inel_main_WANN}] and the SCE expression 
[Eq.~\eqref{eq:SCE_CS_fin}], as a function of $U$ in units of the critical interaction $U_\textrm{c}^{(\nu)}$, for a fixed scattering angle. 
In terms of the renormalized interaction strength, the relative deviations $\Delta_\text{ICS}$ for constant $L$ and different $\nu$ nearly coincide, and they exhibit the same behavior when $L$ is increased.
For $U>U_\textrm{c}^{(\nu)}$ the Mott phase is approached and the deviations become smaller with increasing interaction, whereas for $U\lesssim U_\textrm{c}^{(\nu)}$ the deviations become increasingly pronounced. The non-monotonicity of $\Delta_\text{ICS}$ visible for $L=4$ is due to a crossing of the analytical approximation and the exact numerical data; a finite-size effect which disappears for $L>5$. 
The inset of the lower panel of Fig.~\ref{fig:SCECS_L4_diff_n_v2} reveals that, around the critical point, $\Delta_\text{ICS}$ depends exponentially on the renormalized interaction strength, with an exponent that changes as the transition is crossed and seemingly converges to a system size independent value as $L$ is increased. 

\begin{figure}
\centering
	\includegraphics[width=\figwidth]{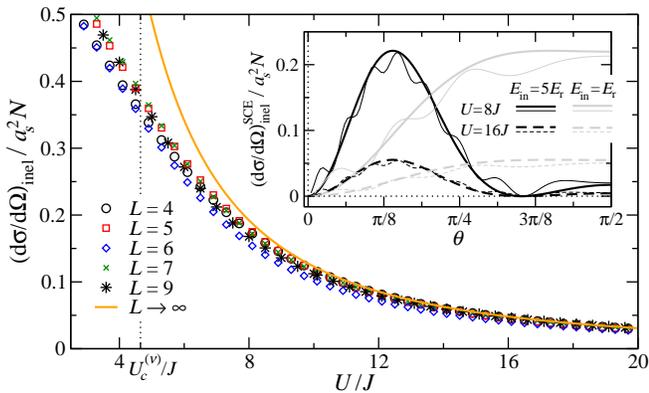}
	\caption{(Color online) Inelastic scattering cross section in the strong-interaction regime for unit filling factor $\nu=1$. The symbols show exact numerical results for different system sizes, while the thick solid line corresponds to the $L\rightarrow\infty$ limit [Eq.~\eqref{eq:SCE_CS_large_L}] of expression \eqref{eq:SCE_CS_fin}, obtained from SCE. The vertical dotted line indicates the critical point of the SF-MI phase transition obtained from third-order SCE (see Appendix \ref{app:SCEformulas}).  
	Relevant parameters are $\Ein=2E_r$, $m=M$ and $\theta = 0.99$. The inset shows the angular dependence of the SCE inelastic cross section for $L=8$ [thin lines, Eq.~\eqref{eq:SCE_CS_fin}] and $L\rightarrow\infty$ [thick lines, Eq.~\eqref{eq:SCE_CS_large_L}], for different values of the incoming energy and interaction strength, as indicated in the legend.}
	\label{fig:SCECS_n1_diffL}
\end{figure}

Here, we focus on the regime where the probe energy is high as compared to that part of the excitation spectrum of the system which contributes significantly to the cross section. Then, the corresponding excited states have comparable weights in the scattering signal [cf. the radicands in Eq.~\eqref{eq:CS_inel_main_WANN}]. Using the excitation gap to first order in $\tilde{J}$, as considered in the inelastic cross section \eqref{eq:SCE_CS_fin}, the high incoming energy condition relevant for strong interaction reads $\Ein\gg U(1-\tilde{J}\EECt{q,\varkappa}{1})$, 
which translates into 
\begin{equation}
 \frac{E_\textrm{in}}{J} \gg \frac{U}{J} + 2\sqrt{4\intFF(\intFF+1)+1}. 
 \label{eq:highEin}
\end{equation}
In this regime the transferred momentum \eqref{eq:kappa_SCE} can be approximated by $\kappa\approx\kappa_\text{el}$, and, most interestingly, 
the inelastic cross section \eqref{eq:SCE_CS_fin} converges to a simple system-size independent expression as $L\rightarrow\infty$ (see Appendix \ref{app:large_L}):
\begin{equation}
	\frac{1}{Na_s^2}\frac{d\sigma}{d\Omega}\bigg|_\textrm{inel}^\textrm{SCE} =  8(\intFF+1)\sin^2 \left(\frac{\kappa_\textrm{el} d}{2}\right)
	\big|W(\kappa_\textrm{el})\big|^2 \left[\frac{J}{U}\right]^2. \label{eq:SCE_CS_large_L} 
\end{equation}
The decay of the exact numerical cross section is compared to the latter analytical formula 
for different system sizes in Fig.~\ref{fig:SCECS_n1_diffL}.
For strong interaction, Eq.~\eqref{eq:SCE_CS_large_L} describes remarkably well the numerical results, which converge very quickly to the $L\rightarrow\infty$ cross section. 
We also note that for a typical choice of $E_\text{in}\sim E_r$, condition \eqref{eq:highEin} translates into a regime for the interaction strength within which the $L\rightarrow\infty$ expression should be valid; for $\nu=1$ it reads $1\ll U/J \ll 148$. 
Equation \eqref{eq:SCE_CS_large_L} also reproduces correctly the overall angular dependence of the inelastic cross section even for not so high interaction, 
as can be seen in the inset of Fig.~\ref{fig:SCECS_n1_diffL}.

The existence of a system-size independent expression for the decay in the regime of strong interaction has a major 
consequence: Since Eq.~\eqref{eq:SCE_CS_large_L} is able to describe the behavior of the inelastic cross section in the Mott insulating regime, we can conclude that, for incoming energies higher than the Mott gap, the cross section is non-zero inside the entire Mott phase in the thermodynamic limit, and does \emph{not} vanish at the critical point. 

\begin{figure}
\centering
	\includegraphics[width=\figwidth]{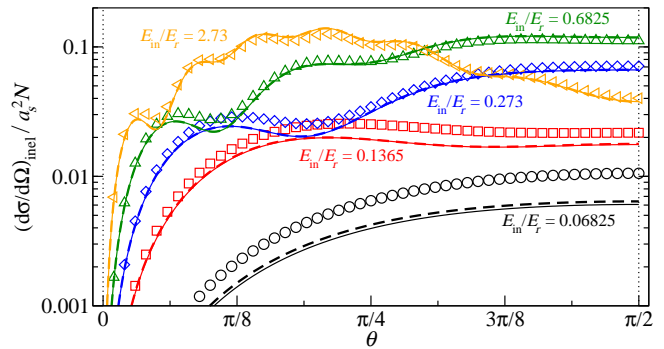}
	\caption{(Color online) Inelastic scattering cross section for different incoming energies $\Ein$ at fixed interaction strength $U/J=10$ and $m=M$, for a system of $L=8$ sites at $\nu=1$, versus scattering angle. Symbols correspond to exact numerical results, while lines show the SCE analytical expression when the energy gap $E_n-E_0$ is evaluated to first order in $\tilde{J}$ [Eq.~\eqref{eq:SCE_CS_fin}, solid lines], and to second order [see Eq.~\eqref{eq:E2}, dashed lines]. We note that the maximum excitation energy in the system is $\max(E_n-E_0)=1.84 E_r$, and the number of contributing states (\% of the total 6435 states) to the exact numerical cross section depends on the value of $\Ein/E_r$: $2.73 (100\%)$, $0.6825 (88.2\%)$, $0.273 (25.3\%)$, $0.1365 (3.4\%)$, $0.06825 (0.4\%)$. The analytical approximation \eqref{eq:SCE_CS_fin} for $L=8$ uses at the most $15$ excited states $\ket{\phi_{q,\varkappa}}$ to describe the cross section.}
	\label{fig:SCECS_diffE0_thetadep}
\end{figure}

Up to second order in $\tilde{J}$, the analytical expression \eqref{eq:SCE_CS_fin} for the inelastic cross section is complete. 
It contains, however, higher incomplete orders in $\tilde{J}$, since the full functional dependence of the weighting factors $\sqrt{1-(E_n-E_0)/E_\text{in}}$, and correspondingly of the transferred momentum $\kappa$, has been kept, where we simply inserted the expansion of the energy gap to leading order in $\tilde{J}$. These incomplete orders are only relevant for low incoming energy. An increasing deviation between Eq.~\eqref{eq:SCE_CS_fin} and the exact results should then be expected as $E_\text{in}$ is reduced  
for a fixed interaction strength, as confirmed in Fig.~\ref{fig:SCECS_diffE0_thetadep}. In this regime, an expansion of the energy gap to second order induces very little improvement, as also seen in Fig.~\ref{fig:SCECS_diffE0_thetadep}. A higher order expansion of the matrix elements \eqref{eq:ME} would be necessary to describe the cross section more accurately for low incoming energy. 
\subsection{Description for all $U/J$: Bogoliubov result and SCE}

\begin{figure}
\centering
	\includegraphics[width=\figwidth]{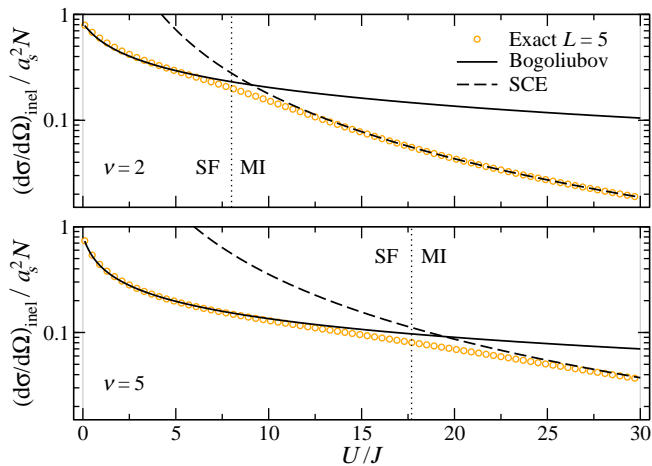}
	\caption{(Color online) Decay of the inelastic scattering cross section (logarithmic scale), as a function of the interaction strength $U/J$, for a system of $L=5$ sites at filling factors $\intFF=2$ (upper panel) and $\intFF=5$ (lower panel). Symbols show exact numerical results, while lines correspond to analytical approximations: the Bogoliubov formula for small condensate depletion, Eq.~(44) in Ref.~\cite{MayRB14} (solid), and the SCE expression \eqref{eq:SCE_CS_fin} (dashed). Vertical dotted lines mark the position of the estimated critical points of the SF-MI phase transition (Appendix \ref{app:SCEformulas}). 
	Relevant parameters are $\Ein=2E_r$, $\theta = 0.99$ and $m=M$.}
	\label{fig:SCEandBogCS_L5B10}
\end{figure}

\begin{figure}
\centering
	\includegraphics[width=\figwidth]{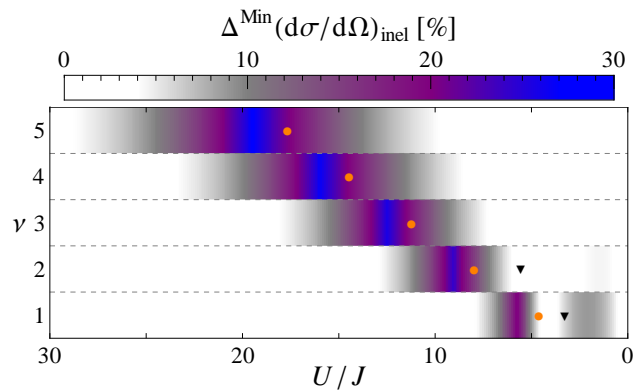}
	\caption{(Color online) Minimum $\Delta^\textrm{Min}\big(d\sigma/d\Omega\big)_\textrm{inel}$ of the relative differences (absolute value) between the exact numerical cross section and 
	the analytical approximations: the SCE result [Eq.~\eqref{eq:SCE_CS_fin}] and the Bogoliubov expression [Eq.~(44) in Ref.~\cite{MayRB14}], as a function of the interaction strength $U/J$, for $L=5$ and different integer bosonic densities. For every value of $U/J$ the relative difference is computed for the SCE expression [see Eq.~\eqref{eq:rel_diff}] and for the Bogoliubov result, and then the minimum of both is taken. 
	Symbols indicate the critical value of the SF-MI phase transition, as predicted from third-order SCE (circles), and from DMRG calculations (triangles, for $\nu=1,2$; data from Ref.~\cite{EjiFG11}). Relevant parameters are $\Ein=2E_r$, $\theta = 0.99$ and $m=M$.}
	\label{fig:RelDiffSCEandBog}
\end{figure}

In Ref.~\cite{MayRB14}, we derived an analytical formula for the inelastic cross section in the regime of weak interaction (small condensate depletion), making use of the Bogoliubov formalism. In combination with the SCE expression obtained in our present contribution, we provide an analytical description of the decay of the inelastic cross section which works remarkably well almost in the entire range of interaction strengths, for any integer filling factor, as shown in Fig.~\ref{fig:SCEandBogCS_L5B10}. 
The shift of the regime of validity of Eq.~\eqref{eq:SCE_CS_fin} to larger $U/J$ for higher $\nu$ is compensated by the extended range of validity of the Bogoliubov approximation at high densities, Eq.~(44) in Ref.~\cite{MayRB14}.
The region where none of the analytical expressions seems to reproduce accurately the exact cross section can be correlated with the position of the SF-MI phase transition (in Fig.~\ref{fig:SCEandBogCS_L5B10}, the critical points estimated from SCE are indicated by vertical dotted lines). 
This can be clearly seen in Fig.~\ref{fig:RelDiffSCEandBog}, 
which shows $\Delta^\textrm{Min}(d\sigma/d\Omega)_\textrm{inel}$, the minimum of the relative differences between the exact inelastic cross section and both analytical formulas, Eq.~(44) in Ref.~\cite{MayRB14} (Bogoliubov) and Eq.~\eqref{eq:SCE_CS_fin} (SCE), as a function of the interaction strength, for different filling factors. 
The approximate expressions deviate from the exact results in the vicinity of the transition, at which perturbative approaches can be expected to fail. Nevertheless, the overall analytical description performs remarkably well. 

\section{Conclusions}
\label{sec-Concl}
We have studied the inelastic cross section of a matter-wave 
scattered from a collection of interacting ultracold bosons in an optical lattice, focusing on the regime of large boson-boson interaction strength. 
We have employed a strong-coupling expansion (SCE) and a site-decoupling mean-field (MF) approach to analytically describe the system eigenstates and energies, from which analytical expressions for the cross section were obtained. The MF approach incorrectly predicts a linear decay of the cross section with increasing interactions, and its  
vanishing as the phase boundary from the SF to the MI regime is crossed. In contrast, SCE predicts a quadratic decay of the
cross section, which vanishes only as $U/J\rightarrow\infty$. 
Moreover, we derived 
an $L\rightarrow\infty$ description of the cross section, which  
reveals that the inelastic scattering signal does not vanish at the SF-MI phase transition in the thermodynamic limit, but decays smoothly throughout the Mott phase. 
These findings are confirmed by exact numerical simulations. 

\begin{figure}
\centering
	\includegraphics[width=\figwidth]{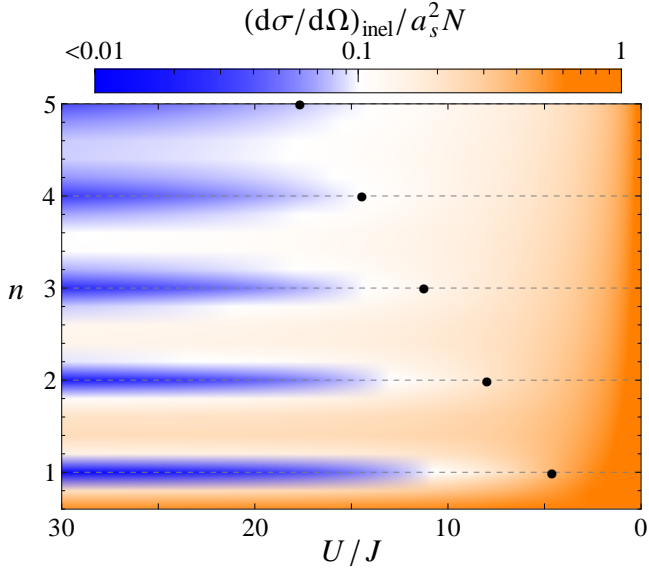}
	\caption{(Color online) Numerically calculated inelastic cross section for a system with $L=5$ versus interaction strength and bosonic density $n\equiv N/L$, where $3\leqslant N\leqslant 25$ (note the inverted abscissa axis). Black circles indicate the critical value of the SF-MI phase transition, as predicted from third-order SCE. Relevant parameters are $\Ein=2E_r$, $\theta = 0.99$ and $m=M$.}
	\label{fig:ExactICSdplot}
\end{figure}

Together with the results previously obtained in the Bogoliubov regime \cite{MayRB14}, we provide a remarkable analytical description of the decay of the scattering signal over almost the entire range of interaction strength, which seems to fail only in the vicinity of the critical point. The question remains open, as to whether or not the cross section bears a quantifiable fingerprint of the phase transition. 
As shown in Fig.~\ref{fig:ExactICSdplot}, the exactly computed inelastic cross section behaves in a clearly distinctive way depending on the phase of the system. For incoming energies larger than the Mott gap, the inelastic cross section will be non-vanishing in the Mott phase, as demonstrated in this work. Nevertheless, the transition in the thermodynamic limit could manifest itself as a non-analyticity at the critical point, i.e.~a discontinuity of one of the higher derivatives of the inelastic cross section. This issue is not revealed within the perturbative approaches pursued in this work, and remains to be clarified. 
\begin{acknowledgments}
We gratefully acknowledge the Deutsche Forschungsgemeinschaft for financial support. 
We thank V.~Shatokhin for insightful discussions and useful comments. 
\end{acknowledgments}

\appendix
\section{Strong-coupling expansion of $\ket{\phi_{q,\varkappa}}$ states and estimation of the fixed-density SF-MI critical point}
\label{app:SCEformulas}
The perturbative expansion of the eigenstates $\ket{\phi_{q,\varkappa}}$ of $\hmu$ which converge as $\tilde{J}\rightarrow 0$ to the one-PH states $\ket{q,\varkappa}$ [Eq.~\eqref{eq:qk_states}] is given in Eqs.~\eqref{eq:ph_states_expansion}. 
The first order correction to the state, $\ket{\phi_{q,\varkappa}^{(1)}}$ is written in Eq.~\eqref{eq:qk_firstorder}, in terms of 
\begin{align}
	\ket{v_{\mathcal{C}}} &= \sum_{\substack{\ESU{}\notin\mathcal{D} \\ \ESU{} \neq \phimi}}\frac{ \braket{ \nvec | \hV | q,\varkappa}  }{ \EEUt{\textrm{ph}} -\EEUt{\nvec} }  \ESU{}, \label{eq:v_C} \\
	\ket{v_{\mathcal{D}}} &= \sum_{\varkappa'\neq\varkappa} 
	\sum_{\ESU{}\notin\mathcal{D}}\frac{\braket{ q,\varkappa'|\hV|\nvec} \braket{\nvec|\hV|q,\varkappa}}
	{\left(\EECt{q,\varkappa}{1} - \EECt{q,\varkappa'}{1}\right) (\EEUt{\textrm{ph}} -\EEUt{\nvec})} \ket{q,\varkappa'}\label{eq:v_D}.
\end{align}
The corrections appearing in the SCE for the energy [Eq.~\eqref{eq:PH_enegy}] up to second order in $\tilde{J}$ are $\EECt{q,\varkappa}{1}$, given in Eq.~\eqref{eq:PH_disp_rel}, and 
\begin{equation}
 \EECt{q,\varkappa}{2}= \sum_{\ESU{}\notin\mathcal{D}}\frac{ |\braket{\nvec | \hV | q,\varkappa }|^2}{\EEUt{\textrm{ph}}-\EEUt{\nvec}},
\end{equation}
 which after a very lengthy calculation can be found to be 
\begin{align}
	&\EECt{q,\varkappa}{2}= 
	-2L\intFF(\intFF+1)+\left(6\intFF^2+6\intFF+1\right) \notag\\
	&\;-\frac{4\intFF(\intFF+1)}{L}\cos(qd)\cos(2\zeta_q -qd)\big[1+(L-1)\cos(2\varkappa d)\big] \notag\\
	&\;+\frac{2\sin^2(\varkappa d)}{3L}\big[2\intFF(\intFF+1)\left(6\cos(qd)-1\right)+ 1 \big] \notag\\
	&\;+\sin(\varkappa d)\sin\big[\varkappa d(L-1)\big]\bigg\{ 4\intFF(\intFF+1)\delta_{q,0} \notag\\
	&\;\quad-\frac{2\intFF(\intFF+2)}{L}\cos\big[\zeta_q (L-2)\big]\notag\\
	&\;\quad-\frac{2\left(\intFF^2-1\right)}{L}\cos\big[\zeta_q (L-2)+2qd\big] \notag\\
	&\;\quad+\frac{4\intFF(\intFF+1)}{L}\cos\big[\zeta_q (L-2)+qd\big] \notag\\
	&\;\qquad\times\big[(1-\delta_{q,0})\big(1+2\cos(qd)\big)-(L-3)\delta_{q,0}\big]\bigg\},
	\label{eq:E2}
\end{align}
valid for $L\geqslant 3$, and where $\zeta_q$ is defined in Eq.~\eqref{eq:zeta_q}.

For the lowest energy fixed-density excitation of $\hmu$, corresponding to $q=0$ ($\zeta_q=0$) and $\varkappa d=\pi/L$, the SCE yields the energy gap [see Eq.~\eqref{eq:Mott_energy_2ndorder} for the ground state energy expansion]
\begin{align}
 \tilde{\Delta}_L =& 1 -2\cos(\pi/L)\sqrt{1+4\nu(\nu+1)} \tilde{J} \notag\\ 
                &+\tilde{J}^2\bigg\{1+ 2\nu(\nu+1)[3-2\cos(2\pi/L)] \notag \\ 
                &+\frac{4}{3L}\sin^2(\pi/L)[5\nu(\nu+1)+2]\bigg\} + \mathcal{O}(\tilde{J}^3),
\end{align}
which is system-size dependent. The critical point of the SF-MI fixed-density phase transition can be estimated from the vanishing of the energy gap in the thermodynamic limit, $\tilde{\Delta}_\infty=0$. The expansion of $\tilde{\Delta}_\infty$ is more easily obtained in a non-number-conserving approach, 
where the fixed-density excitation gap follows by adding the excitation energies from the ground state $\phimi$ to defect states containing one extra particle or hole. 
For these defect states, the SCE approach on a system with periodic boundary conditions provides a size-independent expansion for the gap valid in the thermodynamic limit
\cite{FreM94,FreM96,EjiFGM12,May2014}. From $\tilde{\Delta}_\infty$ up to third order in $\tilde{J}$, the estimation of 
the critical point $\tilde{J}_c^{(\intFF)}$ at integer filling factor $\intFF$ is given by the smallest positive solution of 
\begin{align}
	0=1 -2(2\intFF+1)\tilde{J} + \big[1 + 2 \intFF(\intFF + 1)\big] \tilde{J}^2 + 2\intFF(\intFF^2 + 2) \tilde{J}^3,
\end{align}
which evaluates to $\tilde{J}_c^{(1)}=(\sqrt{7}-2)/3=0.215$ and $\tilde{J}_c^{(2)}=1/8=0.125$. 
The estimates for these two critical points from DMRG calculations are $\tilde{J}_c^{(1)}=0.305$ and $\tilde{J}_c^{(2)}=0.180$ \cite{EjiFG11}.
A comparison of the full phase diagrams $\mu$ vs $U$ obtained from third order SCE and DMRG calculations can be found in Ref.~\cite{May2014}.
\section{Derivation of the $L\rightarrow\infty$ SCE inelastic cross section}
\label{app:large_L}
The limit of large incoming energy, to first order in $\tilde{J}$, is given by the condition
\begin{equation}
	\Ein\gg U(1+\tilde{J}\EECt{q,\varkappa}{1}),\, \forall\ q,\varkappa,
\end{equation}
which from the particle-hole dispersion \eqref{eq:PH_disp_rel} translates into
\begin{equation}
	\frac{\Ein}{J}\gg\frac{U}{J}+2\sqrt{1+4\intFF(\intFF+1)}.
\end{equation}
In this regime, the transferred momentum [Eq.~\eqref{eq:kappa_SCE}] becomes independent of $q$ and $\varkappa$: $\kappa\approx\kappa_\text{el}$, and  
the sum over $\varkappa=\pi j' / Ld$, $j'=1,\ldots,L-1$, in Eq.~\eqref{eq:SCE_CS_fin} can be readily evaluated: 
\begin{equation}
	\sum_{j'=1}^{L-1}\left|M\big[q,\varkappa(j')\big]\right|^2 = 4L\sin^2\left(\frac{qd}{2}\right).
\end{equation}
Thus, the inelastic SCE cross section [Eq.~\eqref{eq:SCE_CS_fin}] for large incoming energies can be approximated by
\begin{align}
	\frac{1}{Na_s^2}\frac{d\sigma}{d\Omega}\bigg|_\textrm{inel}^\textrm{SCE} =& \left[\frac{J}{U}\right]^2 8(\intFF+1)\left|W(\kappa_\textrm{el})\right|^2\notag\\
	 & \times \frac{1}{L^2}\sum_{q}\sin^2\left(\frac{qd}{2}\right) \left|\Sigma(\kappa_\textrm{el}-q)\right|^2.
	\label{eq:SCE_CS_large_Ein}
\end{align}
In the limit of a large lattice, we first note that 
\begin{equation}
	\frac{1}{L}|\Sigma(\kappa_\textrm{el}-q)|^2\underset{L\rightarrow\infty}{\longrightarrow} \frac{2\pi}{d}\sum_{Q}\delta(\kappa_\textrm{el}-q-Q),
\end{equation}
where $Q=2\pi j/d$, $j\in\mathbb{Z}$, is a reciprocal lattice vector. Moreover, the sum over the centre-of-mass quasimomentum $q$ 
can be approximated by an integral:
\begin{equation}
	\frac{1}{L}\sum_{q}\; \underset{L\rightarrow\infty}{\longrightarrow}\;\frac{d}{2\pi} \int_{0}^{2\pi/d}dq.
\end{equation}
Using the two limits above in Eq.~\eqref{eq:SCE_CS_large_Ein} 
leads to the $L\rightarrow\infty$ expression for the inelastic SCE cross section, Eq.~\eqref{eq:SCE_CS_large_L}.

\bibliographystyle{prsty}

\end{document}